\documentclass[conference]{IEEEtran}
\IEEEoverridecommandlockouts

\usepackage{cite}
\usepackage{comment}
\usepackage{amsmath,amssymb,amsfonts} 
\usepackage{algpseudocode}
\usepackage{algorithm}
\usepackage{graphicx} 
\usepackage{float} 
\usepackage{textcomp} 
\usepackage[none]{hyphenat}
\usepackage{eso-pic}
\usepackage{textcomp} 
\usepackage{makecell}
\usepackage{multicol}
\usepackage{multirow}
\usepackage{caption}
\usepackage{tabularx,booktabs}
\usepackage{xcolor}
\usepackage{amsmath,amssymb,amsfonts} 
\usepackage{graphicx} 
\usepackage{float} 
\usepackage{textcomp} 
\usepackage{eso-pic}
\usepackage{xcolor}
\usepackage{makecell}
\usepackage{multicol}
\usepackage{multirow}
\usepackage{caption}
\usepackage{tabularx,booktabs}
\newcolumntype{C}{>{\centering\arraybackslash}X} 
\usepackage{array}
\newcolumntype{P}[1]{>{\centering\arraybackslash}p{#1}}
\usepackage{comment}
\usepackage{csquotes}
\usepackage{lineno,hyperref}

\usepackage{csquotes}
\def\BibTeX{{\rm B\kern-.05em{\sc i\kern-.025em b}\kern-.08em
    T\kern-.1667em\lower.7ex\hbox{E}\kern-.125emX}}
    
\makeatletter
\def\ps@IEEEtitlepagestyle{%
  \def\@oddfoot{\mycopyrightnotice}%
  \def\@evenfoot{}%
}
\def\mycopyrightnotice{%
 {\footnotesize 979-8-3503-9431-3/23/\$31.00 \textcopyright2024 IEEE\hfill}
  \gdef\mycopyrightnotice{}
}

\usepackage[textsize=footnotesize, textwidth=1.2cm]{todonotes}

\usepackage{eso-pic}

\specialcomment{fpal}{\color{red}}{\color{black}}

\begin{document}




\title{IoT-enabled Drowsiness Driver Safety Alert System with Real-Time Monitoring Using Integrated Sensors Technology}

\author{\IEEEauthorblockN{Bakhtiar Muiz\IEEEauthorrefmark{1}, Abdul Hasib\IEEEauthorrefmark{2}, Md. Faishal Ahmed\IEEEauthorrefmark{3}, Abdullah Al Zubaer\IEEEauthorrefmark{4},
Rakib Hossen\IEEEauthorrefmark{4},\\ Mst Deloara Khushi\IEEEauthorrefmark{5}, and Anichur Rahman\IEEEauthorrefmark{6}}

\IEEEauthorblockA{
\textit{Dept. of Internet of Things and Robotics Engineering, Bangabandhu Sheikh Mujibur Rahman Digital University},\\
\textit{Department of Cyber Security Engineering, Bangabandhu Sheikh Mujibur Rahman Digital University, Bangladesh},\\
\textit{Dept. of Computer Science and Engineering, National Institute of Textile Engineering and Research (NITER)},\\
\textit{Constituent Institute of Dhaka University, Savar, Dhaka-1350}\\
bakhtiarmuiz247@gmail.com\IEEEauthorrefmark{1},
sm.abdulhasib.bd@gmail.com\IEEEauthorrefmark{2},
faishalahmed003@gmail.comd\IEEEauthorrefmark{3},
1801027@iot.bdu.ac.bd\IEEEauthorrefmark{4},\\
rakib0001@bdu.ac.bd\IEEEauthorrefmark{5},
khushi001@bdu.ac.bd\IEEEauthorrefmark{6},
anis.mbstu.cse@gmail.com\IEEEauthorrefmark{7}}
}

\maketitle

\begin{abstract}
\boldmath
Significant losses in terms of life and property occur from road traffic accidents, which are often caused by drunk and drowsy drivers. Reducing accidents requires effective detection of alcohol impairment and drowsiness as well as real-time driver monitoring. This paper aims to create an Internet of Things (IoT)--enabled Drowsiness Driver Safety Alert System with Real-Time Monitoring Using Integrated Sensors Technology. The system features an alcohol sensor and an IR sensor for detecting alcohol presence and monitoring driver eye movements, respectively. Upon detecting alcohol, alarms and warning lights are activated, the vehicle speed is progressively reduced, and the motor stops within ten to fifteen seconds if the alcohol presence persists. The IR sensor monitors prolonged eye closure, triggering alerts, or automatic vehicle stoppage to prevent accidents caused by drowsiness. Data from the IR sensor is transmitted to a mobile phone via Bluetooth for real-time monitoring and alerts. By identifying driver alcoholism and drowsiness, this system seeks to reduce accidents and save lives by providing safer transportation.

\end{abstract}
\vspace{2mm}

\begin{IEEEkeywords}
Internet of Things, Drowsiness, Fatigue, Driver Safety Alarm, Sleep Detection, Arduino Uno, IR and Alcohol Detection.
\end{IEEEkeywords}

\section{Introduction}
1.35 million people worldwide lose their lives in traffic accidents each year \cite{cdcGlobalRoad}. This number is 3,700 for a day only. Many of these are due to driver drowsiness and drunk driven, leading to loss of control at high speeds, for that reason accidents happened.
Many rely on alarms to wake up, similar to the need for drowsiness driver safety systems during long journeys. Driver fatigue leads to many fatal accidents due to loss of control at high speeds, causing deaths and severe road damage. For safe transportation, addressing human drowsiness is essential since the number of road accidents caused by sleepy drivers is rising and surpassing other reasons. There are financial, health, and environmental consequences \cite{b1}. 93\% of road accident deaths worldwide take place in low- and middle-income nations. Most nations lose 3\% of their GDP to road accidents \cite{b2}. Bangladesh ranks 106th out of 183 nations in terms of the highest number of road accident-related deaths, according to World Health Organization data. 
Drivers falling asleep is one of the major causes of these accidents. The system assists the driver in receiving an alert for drowsiness. \vspace{1mm}

\par Drunk driving is one of the major causes of accidents as well. Drunk driving is a serious problem that causes many tragic events worldwide. Alcohol and drug use are common among drivers \cite{b3}. Drug use is common among drivers involved in accidents. Young drivers are often more closely linked to drug-related driving when they are under 35 \cite{b4}. Drug consumption reduces alertness. The purpose of our proposed system also includes \enquote{Drunk Driving Detection}. Drivers must ensure they are sober and capable of driving to avoid endangering lives. Drinking may seem harmless, but it becomes dangerous when operating a large vehicle, potentially causing tragic accidents involving innocent people. A drunk driver may operate a vehicle four times in forty-eight minutes on a median before being pulled over. Moreover, the consequences of drunk driving affect a larger number of people than one may think, as it costs each adult in our nation around \$500 annually\cite{a1}.\vspace{1mm}

\textcolor{black}{While many studies detect drowsiness and alcohol impairment, none combine eye blink monitoring, alcohol detection, Bluetooth alerts, vibration notifications, and automatic vehicle stoppage. For example, \cite{Ahmed} integrates eye blink detection but lacks full integration of Bluetooth and vehicle stoppage. This research uniquely combines all these elements, making it a valuable contribution to driver safety.}

This study, \enquote{IoT-enabled Drowsiness Driver Safety Alert System with Real-Time Monitoring} can stop these problems and save a lot of lives as well as prevent accidents. This proposed system has two main sections: driver alertness monitoring and alcohol detection, integrated with mobile communication via Bluetooth. When the IR sensor detects closed eyes, it activates vibration, an alarm, and a red warning light after two seconds. If the eyes remain closed, the vehicle slows and stops within 10-15 seconds, controlled by an Arduino Uno. Alerts are sent to the driver’s mobile. For alcohol detection, the sensor triggers an alarm and red light for 20 seconds. If alcohol is detected again, the vehicle slows and stops within 10-15 seconds. Commands and alerts are wirelessly transmitted to the driver’s mobile via Bluetooth, ensuring seamless communication and safety protocol execution \cite{rahman2022sdn, islam2021blockchain}.\vspace{2mm}
 
In this paper, we design a system drivers can use to reduce road accidents while sleeping. The significant contributions of this paper are summarized below:\vspace{2mm}

 \begin{itemize}
    \item This monitor driver drowsiness through IR sensors, triggering alarms, vibrations, and visual warnings to prevent accidents.

    \item The authors use onboard alcohol sensors to detect alcohol presence, activate warnings, and safely slow and stop the vehicle if necessary.
    
    \item The presented system utilizes Bluetooth for real-time alerts and commands to the driver's mobile device, ensuring efficient execution of safety protocols.
    
    \item Additionally, this study developed system gradually reduces vehicle speed and stops the motor within 10-15 seconds if unsafe conditions are detected, ensuring driver and passenger safety.

\end{itemize} \vspace{1mm}

 \textbf{Organization:} This study's structure is set up as follows: In Section II, we evaluate the literature and talk about previous studies and current technologies that are relevant to our investigation. Section III details the methodology, outlining the design and implementation of our IoT-enabled Drowsiness Driver Safety Alert System. Moreover, in Section IV, we present our system's experimental results and performance metrics. Section V delves into the discussion and limitations, identifying the constraints of our work. 
 Finally, Section VI encapsulates the findings and implications of our research, offering suggestions for future directions.

\section{\textcolor{black}{Related Works}}
Driver fatigue and drowsiness are the major causes of road accidents globally, and hence, various drowsiness technologies and interventions are being developed. Tiwari et al. \cite{tiwari2019iot} solved this problem by using Raspberry Pi 3 B to monitor the driver’s eye condition and health metrics and send alerts in time to ensure safety. Yang et al. \cite{l5} studied the sleep-wake mechanism and used Bayesian networks in a simulated driving environment to detect drowsiness. Sathya et al. \cite{sathya2020iot} proposed an IoT-based system that combines camera-based drowsiness detection with IR sensors and GPS location sharing for safety. Li et al. \cite{l3} used SVM, a probabilistic model, and a smartwatch-integrated EEG device, to monitor drowsiness continuously. Biswal et al. \cite{biswal2021iot} used Raspberry Pi and Pi camera modules to analyze eye blinks through Video Stream Processing (VSP) techniques.

\begin{table}[h]
\scriptsize
\tiny
\caption{Comparison of IoT-Enabled Systems for Drowsiness Driver Safety Alert}
\centering
\scriptsize
\begin{tabular}{|p{2cm}|p{2.5cm}|p{2.5cm}|p{2.5cm}|}
\hline
\textbf{Study} & \textbf{Technologies Used} & \textbf{Key Features} \\ \hline
Tiwari et al. (2019) \cite{tiwari2019iot} & Raspberry Pi 3 B, eye monitoring, health metrics & Timely alerts \\ \hline
Yang et al. (2020) \cite{l5} & Sensors, data analysis & Rule-based task analysis \\ \hline
Sathya et al. (2020) \cite{sathya2020iot} & Cameras, sensors, ML algorithms & Real-time monitoring with ML-based alerts \\ \hline
Li et al. (2020) \cite{l3} & Wearable EEG, SVM & Probabilistic drowsiness estimation \\ \hline
Biswal et al. (2021) \cite{biswal2021iot} & Raspberry Pi 3, Pi camera, VSP & Eye blink analysis (EAR) \\ \hline
Eleyan et al. (2021) \cite{eleyan2021road} & Image processing & Eye blink and facial movement monitoring \\ \hline
Civik et al. (2023) \cite{civik2023real} & Deep learning CNN & Analyze driver's facial features \\ \hline
Himaswi et al. (2023) \cite{10090863} & Raspberry Pi 4, Pi camera, Bluetooth headset, SVM, Histogram of Gradients & Analyze facial features, eye movements, EAR \\ \hline
Saranya et al. (2023) \cite{10084081} & GMM clustering, KNN, SVM, Naive Bayes, Haar-AdaBoost, Fuzzy logic & Steering wheel monitoring, auto shutdown \\ \hline
Jadhav et al. (2023) \cite{jadhav2023anti} & Arduino, face detection & Face detection \\ \hline
Das et al. (2024) \cite{das2024iot} & CNN-LSTM, U-Net segmentation & Facial movement analysis, CNN and LSTM integration \\ \hline
Proposed System & Arduino Uno, alcohol sensor, IR sensor, Bluetooth module, alarm system, vibration system, motor control relay & Advanced sensor integration, speed adjustment, wireless alerts, dual alert mechanisms \\ \hline
\end{tabular}

\label{tab:comparison}
\end{table}

Eleyan et al. \cite{eleyan2021road} mentioned that image processing is needed to detect driver fatigue but with the challenges of false alarms due to varying light conditions. Civik et al. \cite{civik2023real} developed a system that uses real-time video data to track facial and eye movements but doesn’t consider body posture. Himaswi et al. \cite{10090863} used Deep Learning and IoT to analyze eye state and yawning but had issues with light conditions and camera performance. Saranya et al. \cite{10084081} introduced a comprehensive detection method using IoT and OpenCV but had problems with storage space and corruption in portable PCs. Jadhav et al. \cite{jadhav2023anti} used Arduino and face detection to prevent accidents due to fatigue. And Das et al. \cite{das2024iot} proposed a system based on CNN-LSTM and U-Net architectures to detect drowsiness but had challenges with skin tones, diverse backgrounds and head movements.\vspace{2mm}

This system is designed to increase driver safety by combining multiple features with Arduino Uno. It has an alcohol sensor for intoxication detection, IR sensors for driver fatigue monitoring, and a Bluetooth module for wireless alerts. It also has an alarm and vibration system for dual alerts and a motor control relay to adjust vehicle speed based on driver condition. This whole setup combines sensor integration, speed adjustment, and wireless alerts to prevent accidents due to drowsiness or intoxication. Table \ref{tab:comparison} shows the differences and similarities with other studies and how our system integrates multiple sensors and communication technologies to increase driver safety.

\section{Proposed Methodology for IoT-enabled Anti-Sleep Driver Safety Alert System}

\begin{figure}[htb!]
\centerline{\includegraphics[height=7cm, width=8cm]{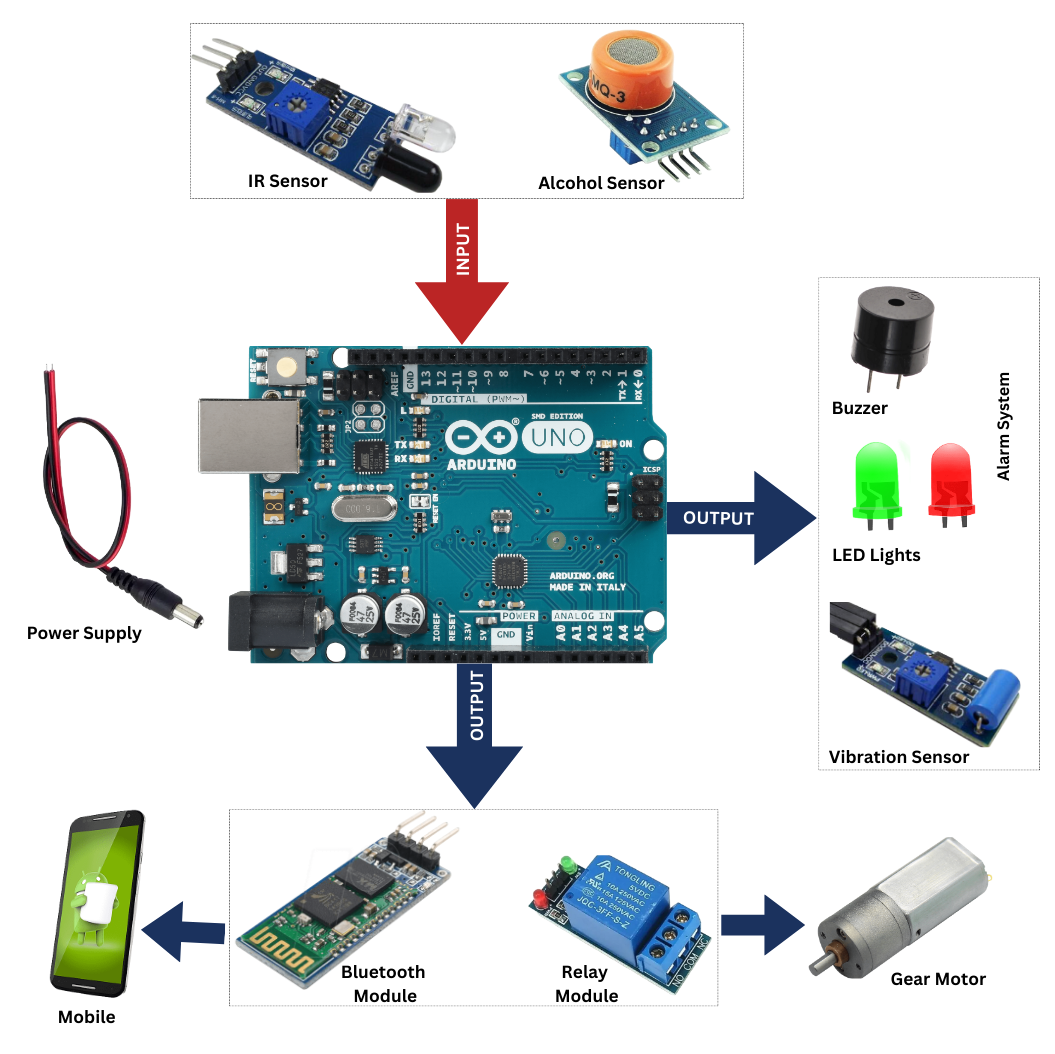}}
\caption{Proposed Block Diagram for Controlling Each Connection}
\label{fig:f1}
\end{figure}

In this system, we have developed a driver alert system for drowsiness by enhancing its RF and Arduino circuits for vehicle integration. This system uses an Alcohol sensor to detect alcohol presence and an IR sensor to monitor the driver's eye movements. The Arduino Uno processes sensor data in real-time, activating alarms, controlling the vehicle's motor, and facilitating wireless communication via a Bluetooth module with the driver's mobile device. Fig. \ref{fig:f1} illustrates the integrated components and their interactions in the proposed block diagram for controlling each connection.

\subsection{Collecting of Hardware Requirements}
This system requires several essential hardware components to function effectively.  The main components used in our project, along with their specifications and functions, are outlined below:

\subsubsection{IR Sensor}
An Infrared Sensor (IR Sensor) detects light in the wavelength range of 780 nm to 50 µm is commonly used for motion detection. These sensors sense changes in heat radiation, detecting human movement within a specified angle range. IR sensors are widely used for their cost-effectiveness and mass production capabilities \cite{rahman2023towards}.
\textbf{Function:}
Monitors driver alertness; triggers alarms and alerts if eyes are closed for prolonged periods. \textcolor{black}{IR sensor is integrated into glasses worn by the driver, continuously monitoring eye movements to detect prolonged eye closure, a sign of drowsiness. The sensor's reliability was validated through controlled experiments where hand movements were used to simulate eye closure, ensuring the sensor's functionality and response time. Once the glasses are worn by the driver, the sensor effectively tracks eye movements, accurately detecting drowsiness through eye closure.}

\subsubsection{Alcohol Sensor}
The MQ3 sensor, a Metal Oxide Semiconductor (MOS) type, detects alcohol by measuring changes in resistance when exposed to it, functioning like a chemiresistor. Operating on 5V DC with a power consumption of approximately 800mW, it can detect alcohol concentrations ranging from 25 to 500 ppm \cite{lastminuteengineers_mq3}. \textbf{Function:}
Trigger warnings and alarms in the event that alcohol is found near the driver.
\textcolor{black}{Alcohol sensor (MQ3) is installed in the driver's seat area, specifically designed to measure the driver’s alcohol levels in the immediate environment. This ensures accurate detection of alcohol impairment. The sensor continuously monitors for alcohol presence and triggers alerts if detected. Its performance has been validated through exposure to controlled alcohol concentrations, confirming its reliability in real-world driving conditions.}

\subsubsection{Arduino UNO}
The Arduino UNO is a microcontroller board based on the ATmega328P. It features a 16 MHz ceramic resonator, 6 analog inputs, 14 digital I/O pins (6 of which can be used for PWM), a USB port, power connector, ICSP header, and reset button—all essentials for the microcontroller's operation \cite{rahman2023icn}. \textbf{Function:}
Controls overall system operations, including sensor data processing, alarm triggering, and motor control.
 
\subsubsection{Bluetooth Module}
The HC-06 Bluetooth module facilitates short-range wireless data transfer using Bluetooth 2.0 with frequency hopping spread spectrum (FHSS). It supports speeds up to 2.1 Mb/s and operates within the 2.402 GHz to 2.480 GHz frequency range for reliable communication \cite{components101_hc06}. \textbf{Function:}
Facilitates communication between the system and the driver's mobile device for real-time alerts and notifications.

\subsubsection{Relay Module}
 A relay is an electrical switch used to control devices and systems operating at higher voltages. They can control electrical loads of both AC and DC types within specified limits. Relay modules come in various voltage ratings, such as 3.2V or 5V for low-power applications and 12V or 24V for heavier-duty systems \cite{geya_relay_module}. \textbf{Function:}
Controls motor operation based on system commands, such as stopping or reducing speed in response to detected risks.
 
\subsubsection{Red and Green LEDs}
The 5mm red and green LEDs emit light with a brightness range of 12-30 millicandela (mcd). They operate at a maximum current of 20 mA with a forward voltage of 1.8V and can handle up to 5V in reverse voltage \cite{robodocbd_led}. \textbf{Function:}
 Red light indicates warnings (e.g., closed eyes, alcohol detection), while green light confirms open eyes status.
 
\subsubsection{Alarm System(Buzzer)}
A buzzer converts electrical signals into sound and can be electromechanical, piezoelectric, or mechanical. Operating on DC voltage, buzzers are used in timers, alarms, printers, computers, and other devices needing audible alerts or notifications \cite{elprocus-buzzer}. \textbf{Function:}
The buzzer functions to alert the driver when detecting closed eyes or alcohol presence, ensuring immediate response to prevent accidents and promote safe driving practices.

\subsubsection{Mobile Device (Smartphone)}
Android device.
\textbf{Function:}
Receives alerts and notifications via Bluetooth; allows user interaction and intervention.
This comprehensive hardware setup enables our system to autonomously monitor driver alertness and detect alcohol presence, ensuring enhanced safety measures in various operational environments \cite{rahman2023impacts}.

\subsection{Arrangement of Software Requirements}
\subsubsection{Arduino IDE}
The Arduino IDE supports Linux, Mac OS X, and Windows, facilitating programming in C and C++. Programs, known as sketches, are created and uploaded to Arduino boards via the IDE for execution \cite{arduino_ide_v1_tutorial}.

\subsubsection{Mobile Application}
'Bluetooth Terminal' is an Android app that connects smartphones to microcontrollers and Arduinos via Bluetooth. It supports Bluetooth LE devices (BBC mini, HM-10) and Bluetooth Classic devices (HC-05, HC-06, Raspberry Pi 3), as well as custom profiles like TI CC2640 and Microchip BM70/71 \cite{cnet_serial_bluetooth_terminal}.

\subsection{Analysis of System Design}
Our IoT-enabled Drowsiness Driver Safety Alert system integrates two critical functions: Alcohol Detection and Driver Alertness Monitoring, complemented by Bluetooth-enabled communication with the driver's mobile device. This setup ensures proactive safety measures against driver fatigue and alcohol impairment. Algorithm \ref{alg:pseudocode} outlines the system's operations, demonstrating its effective technology designed to enhance driver safety and prevent accidents on the road.

\begin{algorithm}
\scriptsize
\caption{IoT-enabled Drowsiness Driver Safety Alert System}
\label{alg:pseudocode}
\begin{algorithmic}[1]
\State \textbf{Initialize:} $\{A, I, B\}$ 
\Procedure{Monitor\_Driver\_State}{}
    \While{$V = 1$} 
        \State $A_D \gets f(A)$ \Comment{Check alcohol sensor}
        \If{$A_D = 1$} 
            \State $\mathbb{A} \gets 1, \mathbb{R} \gets 1$ \Comment{Activate alarm and red light}
            \State \textbf{Wait} $T_A$ \Comment{Wait 20 seconds for recheck}
            \If{$A_D = 1$} 
                \State $\Delta v \gets 1$ \Comment{Slow down}
                \State $\Delta v \gets 0$ \Comment{Stop}
            \EndIf
        \EndIf
        \State $E_C \gets f(I)$ \Comment{Check IR sensor for eye closure}
        \If{$E_C = 1$} 
            \State $B \gets 1$ \Comment{Send alert via Bluetooth}
            \State \textbf{Wait} $T_E$ \Comment{Wait 2 seconds for recheck}
            \If{$E_C = 1$} 
                \State $\mathbb{V} \gets 1, \mathbb{A} \gets 1, \mathbb{R} \gets 1$ \Comment{Activate vibration, alarm, and red light}
                \State $B \gets 1$ \Comment{Send another Bluetooth alert}
                \State \textbf{Wait} $T_E$ \Comment{Wait 2 more seconds for recheck}
                \If{$E_C = 1$} 
                    \State $\Delta v \gets 1$ \Comment{Slow down}
                    \State $\Delta v \gets 0$ \Comment{Stop}
                \EndIf
            \EndIf
        \EndIf
    \EndWhile
\EndProcedure
\end{algorithmic}
\end{algorithm}

When the alcohol sensor ($A$) detects alcohol, the system activates an alarm (\texttt{Alarm\_A}) and illuminates a red warning light across the vehicle for 20 seconds. The system then rechecks the alcohol presence. If alcohol persists ($A_{\text{persist}}$), the system initiates a gradual reduction in vehicle speed ($v_{\text{slow}}$) and eventually stops the motor within 10 to 15 seconds ($t_{\text{stop}}$):
\begin{align}
v_{\text{slow}} &= f(A_{\text{persist}}) \tag{1} \\
t_{\text{stop}} &= g(A_{\text{persist}}) \tag{2}
\end{align}

When the IR sensor detects closed eyes ($E_{\text{closed}}$), it first verifies for 2 seconds ($t_{\text{verify}}$) before activating vibrations (\texttt{Vibrations}), sounding an alarm (\texttt{Alarm\_E}), and lighting up a red warning (\texttt{Warning\_E}). If the eyes remain closed, the system implements a similar procedure to slow down the vehicle and stop the motor within 10 to 15 seconds:
\begin{align}
t_{\text{verify}} &= 2 \text{ seconds} \tag{3} \\
v_{\text{slow}} &= h(E_{\text{closed}}) \tag{4} \\
t_{\text{stop}} &= i(E_{\text{closed}}) \tag{5}
\end{align}

Alerts and commands are promptly relayed to the driver's mobile device via a Bluetooth module, which interfaces with the Arduino Uno for executing critical safety protocols. Bluetooth communication parameters such as data transfer rates (\texttt{data rate}), latency (\texttt{delay}), and error correction (\texttt{EC}) ensure efficient and reliable transmission:
\begin{align}
\texttt{data rate} &= \frac{\text{data transmitted}}{\text{time}} \tag{6} \\
\texttt{delay} &= \text{time}_{\text{end}} - \text{time}_{\text{start}} \tag{7} \\
\texttt{EC} &= \frac{\text{errors corrected}}{\text{total errors}} \tag{8}
\end{align}

This comprehensive system architecture effectively enhances driver safety by integrating real-time sensor data, precise control mechanisms, and seamless wireless communication with the driver.

\subsection{Working Procedure of System Model}
\begin{figure}[h]
\centerline{\includegraphics[height=8cm, width=8cm]{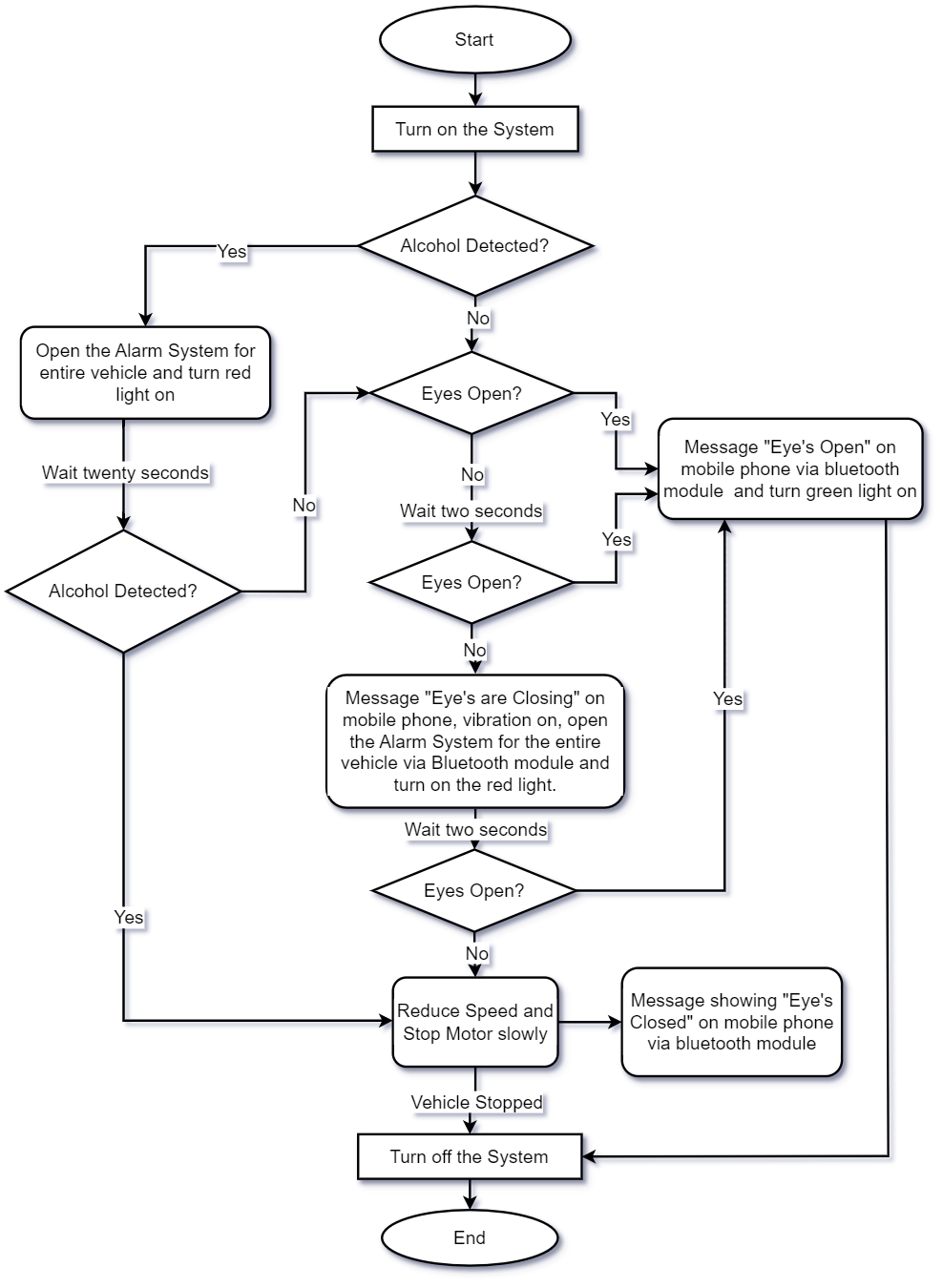}}
\caption{Workflow Diagram of Proposed System}
\label{fig:f2}
\end{figure}

We have designed a system focused on ensuring driver safety by monitoring alertness and detecting alcohol presence automatically. Fig. \ref{fig:f2} illustrates how the system uses advanced sensors and controls to manage fatigue and alcohol presence, enhancing road safety and driver well-being.
 
\begin{figure}[h]
\centerline{\includegraphics[scale=0.23]{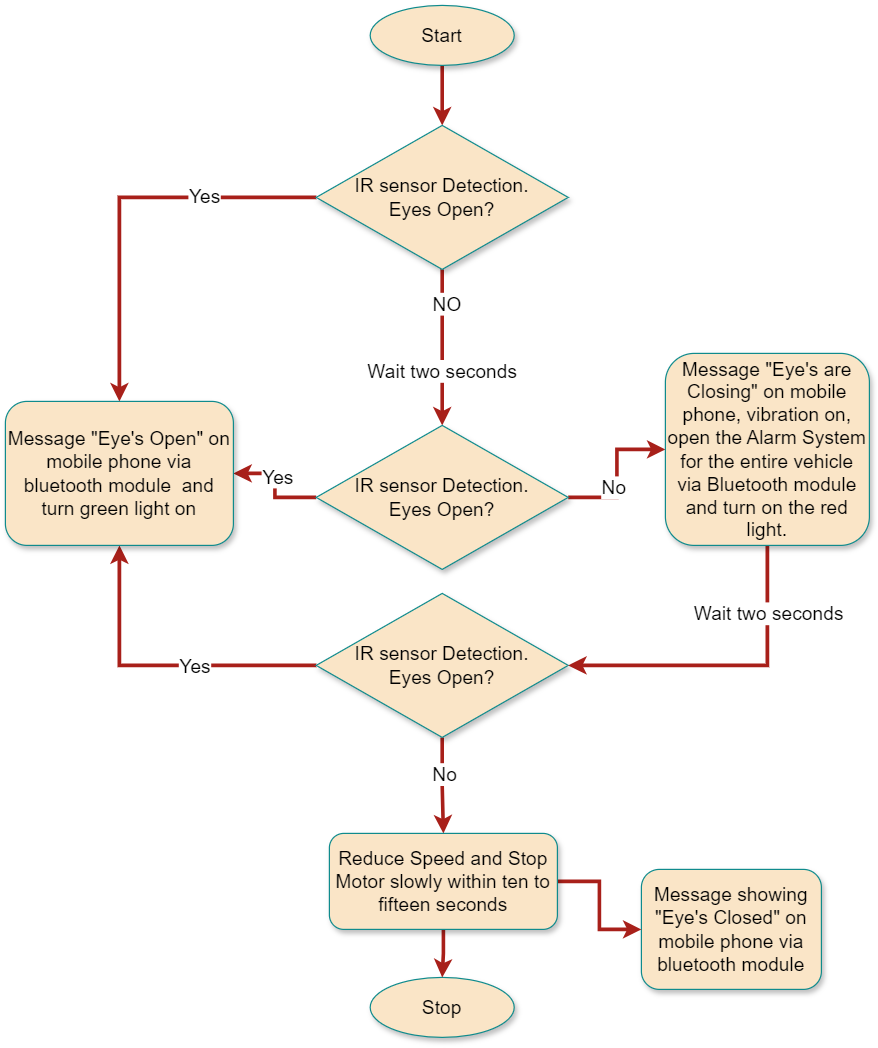}}
\caption{Workflow of IR Sensor}
\label{fig:f3}
\end{figure}

The system includes an Alcohol sensor. When alcohol is detected, it triggers an alarm and lights up a warning for twenty seconds, signaling potential impairment. The system rechecks for alcohol and can adjust vehicle speed or stop it completely for safety. Moreover, this system also uses an infrared (IR) sensor to check if the driver's eyes are closed. If closed eyes are detected, the system waits briefly and then activates an alarm, red warning light, and vibrations to alert the driver. If the eyes remain closed, indicating possible drowsiness, the system gradually slows down the vehicle and can stop it using an Arduino Uno-connected relay. This swift response is crucial for preventing accidents, as shown in Fig. \ref{fig:f3}. Using Bluetooth, the system sends real-time alerts and updates to the driver's mobile device. This helps the driver respond promptly to alerts and stay informed about their driving conditions.

\section{Result Analysis and Performance Measurement}

\subsection{Performance Analysis of Alcohol and IR sensor}
\begin{figure}[H]
\centerline{\includegraphics[height=4cm, width=8cm]{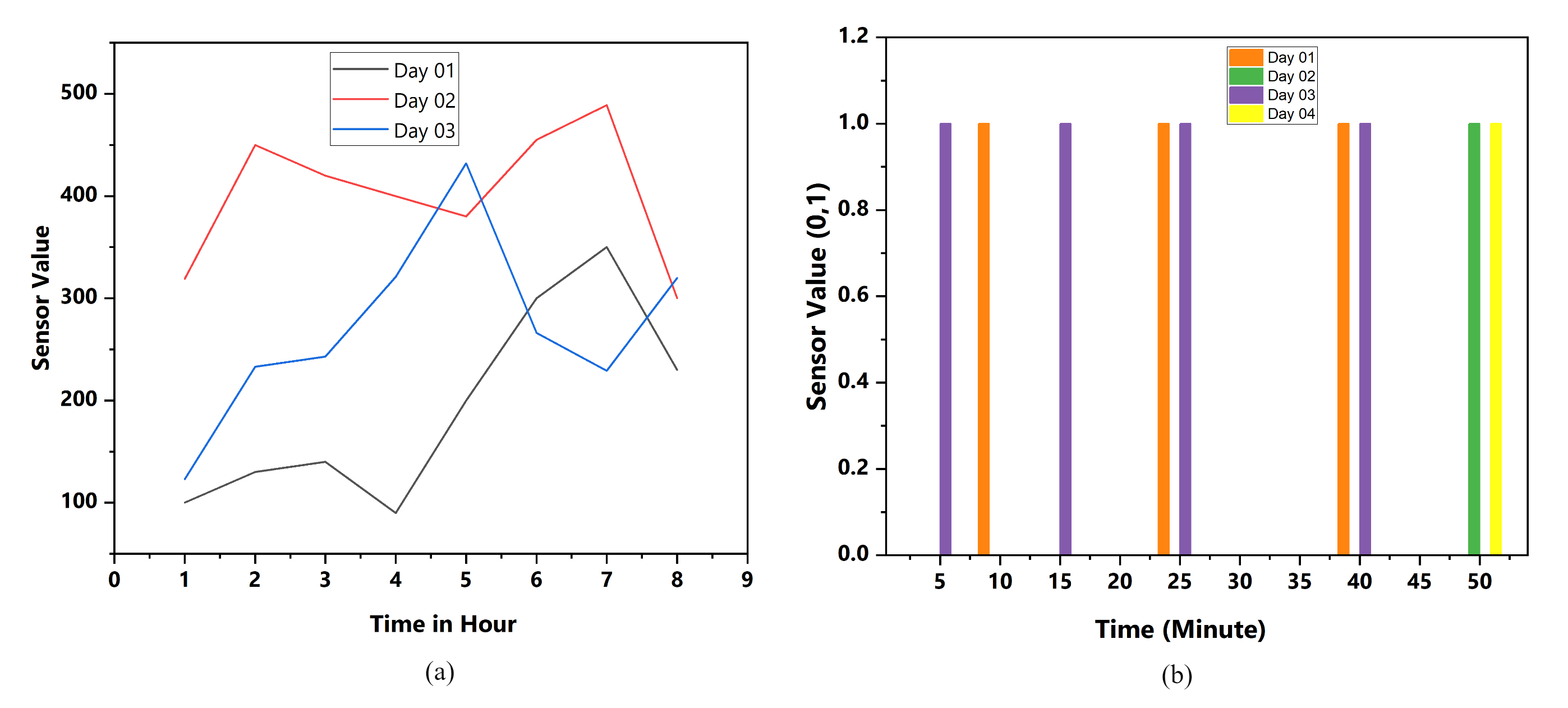}}
\caption{(a) Alcohol Sensor Value Monitoring, (b) IR Sensor Value Monitoring}
\label{fig:f9}
\end{figure}

Fig. \ref{fig:f9}(a) illustrates the performance analysis of an alcohol sensor used to determine driver intoxication over three days, with data collected for eight hours each day. The sensor indicated the presence of alcohol when readings exceeded a threshold of 400. The previous study did not have real-time monitoring accuracy, whereas our system surpassed it.

On the other hand, Fig. \ref{fig:f9}(b) presents a performance analysis of an IR sensor used to monitor a driver's eye movements over a period of four days. Compared to relevant studies, our system more efficiently identified eye movements.

\subsection{Comparison with IoT and Non-IoT System}

\begin{figure}[h]
\centerline{\includegraphics[height=2.90cm, width=8.5cm]{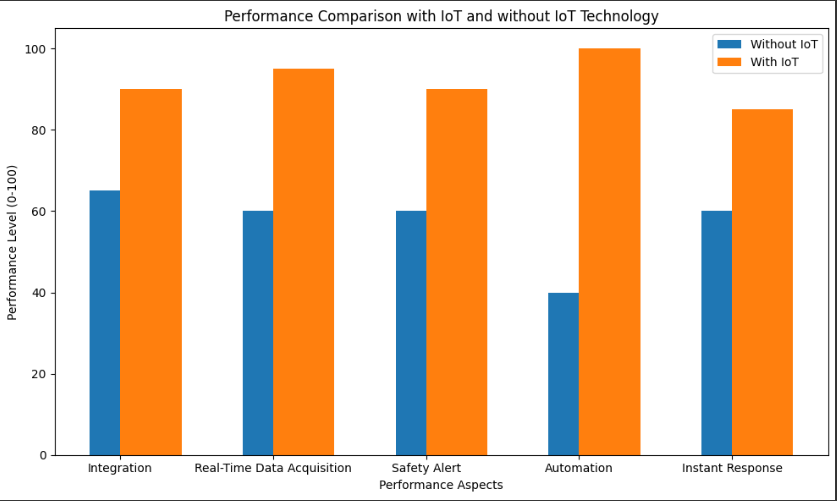}}
\caption{Comparison of the Proposed System with the Non-IoT System}
\label{fig:f8}  
\end{figure}

Fig. \ref{fig:f8} compares key performance metrics—Integration, Real-Time Data Acquisition, Safety Alert, Automation, and Instant Response—between systems utilizing IoT technology and those without it. Each metric is rated from 0 to 100. The graph clearly illustrates IoT's significant enhancements: Integration and Real-Time Data Acquisition score 90 and 95, respectively, with IoT, compared to approximately 65 and 60 without it. Safety Alert improves from 60 to 90, Automation jumps from about 40 to a perfect 100, and Instant Response increases from around 60 to 85 with IoT. \textcolor{black}{The non-IoT system data was gathered through literature, driver interviews, and prior studies without IoT integration. For the IoT system, a driver used glasses with an IR sensor to monitor eye blinks and detect drowsiness. Alerts were triggered based on prolonged eye closures, with reaction times recorded. These were compared to the non-IoT system.} This visual representation highlights the substantial benefits of adopting IoT to improve these critical performance aspects \cite{rahman2024blocksd}.

\subsection{Real Time Analysis}
It enhanced a system to help keep drivers awake and alert by improving its RF and Arduino circuits. Designed for use in vehicles, the system includes an MQ3 sensor to detect alcohol and an IR sensor to monitor the driver’s eye movements. An Arduino Uno processes the sensor data, activates alarms, controls the vehicle’s motor, and communicates via a Bluetooth module. If the MQ3 sensor detects alcohol, the system triggers an alarm and a red warning light for twenty seconds throughout the vehicle. After this period, the system rechecks for alcohol. If alcohol is detected again, the system slows down and stops the vehicle to prevent accidents. The system employs the IR sensor to monitor the driver’s attention if alcohol is not detected. Every two seconds, the IR sensor checks the driver’s eyes to see whether they are open or closed. \textcolor{black}{Figure \ref{fig:rta} illustrates a proof-of-concept model using an Arduino board on a car to simulate real-world scenarios. Driver response was monitored via IR sensors in glasses to detect eye blinks and prolonged closures, simulating drowsiness. In Figure \ref{fig:rta}(c), measurements were taken manually to test the alert system. Drivers were interviewed about their situations and environments, providing insights into real-world responses to drowsiness detection systems without IoT components \cite{faisal2022sdn}.}

\subsubsection{Normal Driving}
When the driver is driving normally without any signs of drowsiness, the system sends a notification to the driver’s phone indicating that their eyes are open and they are alert, as depicted in Fig. \ref{fig:rta}(a).

\subsubsection{Sleepy Driving}
If the IR sensor detects that the driver is starting to feel sleepy, it activates an IR LED, a red light, and starts vibrations. The system constantly monitors the driver’s eyes and sends regular notifications to the driver’s phone if their eyes close. Fig. \ref{fig:rta}(b) illustrates this process. Additionally, an alarm system is activated throughout the vehicle to alert all occupants. This serves as an early alert, encouraging the driver to take a break or pull over safely \cite{islam2020sdot}.

\subsubsection{About to Sleep}
\begin{figure}[!htb]
\centerline{\includegraphics[height=6cm, width=9cm]{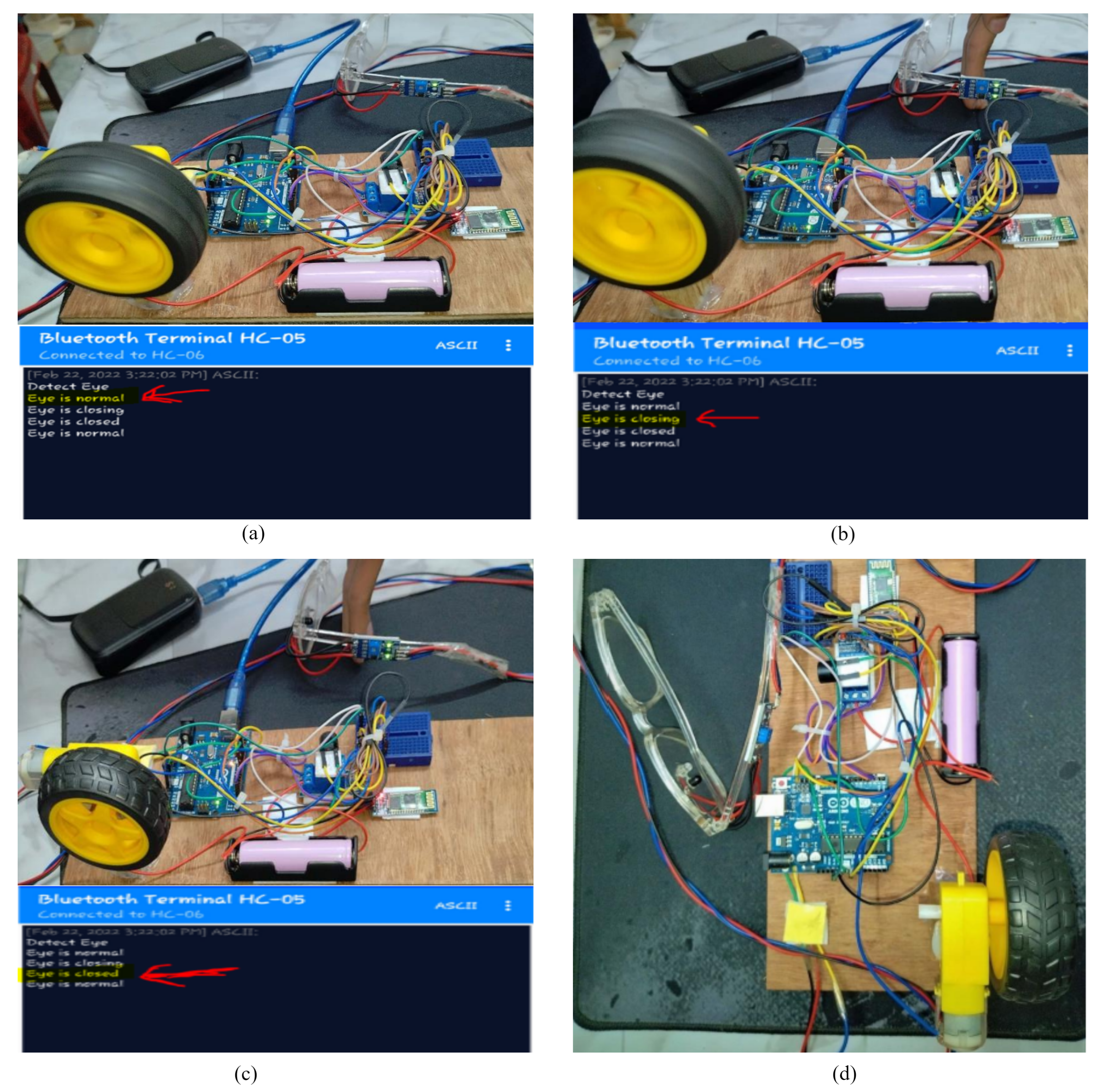}}
\caption{(a) Normal Driving \& Application Interface, (b) Sleepy Driving \& Application Interface, (c) About to Sleep \& Application Interface, (d) Partial Prototype of the System}
\label{fig:rta}  
\end{figure}

On the other hand, as shown in Fig. \ref{fig:rta}(c), when the IR sensor detects that the driver is about to fall asleep, the system sends an urgent notification to the driver’s phone indicating that their eyes are closed. The system intervenes by gradually reducing the vehicle’s speed and ultimately stopping it within 10 to 15 seconds, using a relay connected to the Arduino Uno. The purpose of this warning is to arouse the driver and advise them that they should stop driving immediately in order to prevent other accidents from happening! It will also chime, vibrate, and illuminate the red warning light in response to eye closure. If the driver continues to close his or her eyes, then the system will slow down and finally stop by a relay of the Arduino Uno board, which is interfaced. Fig. Figure \ref{fig:rta}(d) shows an adumbrative representation of the system. The Bluetooth module sends commands and alerts to the driver's mobile phone, enabling quick communication of messages as well as warning alarms in time \cite{rahman2022enhanced}. \vspace{2mm}

In summary, our system uses the MQ3 sensor, IR sensor, and Bluetooth module with precise controls to protect drivers from fatigue and alcohol. The system represents a major strength in the car protection era, with the aid of enhancing street protection and motivating force safety through monitoring and intervention.

\section{Discussion and Limitations}
The IoT-enabled Anti-Sleep Driver Safety Alert System combines sensors, specific controls, and real-time monitoring to noticeably enhance street safety. It makes use of infrared (IR) sensors to detect whilst drivers near their eyes and an MQ3 sensor to discover alcohol tiers, tracking the driver's alertness and impairment. Powered by way of an Arduino Uno, the machine strategies sensor facts to prompt alarms and vibrations or modify vehicle pace to hold drivers secure. Bluetooth conversation sends on-the-spot alerts to a mobile tool, allowing brief responses from the driving force or tracking offerings. 

\subsection{Applications}
The system offers significant benefits in enhancing road safety across various applications:
\begin{itemize}
\item Commercial and Public Transportation: Enhances safety for professional drivers in taxis, buses, and trucks and improves passenger safety in public transportation by preventing accidents due to driver fatigue or impairment.

\item Private Vehicles and Long-Distance Travel: Provides added safety during long-distance travel or late-night driving for individual drivers and passengers.

\item Fleet Management Services: Monitors driver behavior to identify risks and improve overall fleet safety.

\item Emergency Response and Law Enforcement Vehicles: Ensures driver alertness during critical situations, reducing response times and improving outcomes.

\end{itemize}

\subsection{Limitations and Future Improvements}
It has run several tests and found some limitations.
\begin{itemize}

\item Limitations:
\begin{itemize}
\item Sensor Accuracy and Reliability: IR and MQ3 sensors are affected by environmental factors like light and temperature.

\item User Adaptation and Acceptance: The system relies on drivers trusting and responding to alerts.
\end{itemize}

\vspace{2mm}
\item Future Improvements:
\begin{itemize}
\item Better Sensor Accuracy: Calibrate better and use higher accuracy sensors.

\item Better UI \& Installation: Clearer, more intuitive alerts with visual, audible, and haptic feedback. 

\end{itemize}
\end{itemize}

This will help reduce driver fatigue and alcohol impairment and overall road safety and driver happiness.

\section{Conclusion}
Integrating integrated sensor technologies with an IoT enabled anti-sleep driver safety warning system can help save a large number of traffic accidents caused by drunk drivers and sleepers, meanwhile the impact on-revision to reduce-threat recommendation may be effectively reduced due to real-time monitoring. The combination of the infrared sensor with an alcohol sensor can cause the system to detect a driver who is impaired by alcohol and track his or her eye movements. Alarms and warning lights that activate instantly, a vehicle's speed that is gradually lowered, and the potential for a motor shutdown all guarantee prompt action to prevent accidents. Continuous observation and prompt response to any dangers are made possible via Bluetooth real-time data streaming to a mobile device. This method solves the big problems of both drunk and tired driving, making travel a lot safer. They seek to reduce crashes and save lives with their project to make transportation better for everyone. In the future, larger-scale instruments that can detect more features, such as machine learning-based accurate face recognition with a camera, will be useful for this study, and we plan to build a whole model for detecting drunkenness and sleepiness.

\ifCLASSOPTIONcaptionsoff
  \newpage
\fi

\bibliographystyle{IEEEtran}

\bibliography{sample}

\end{document}